\documentclass{article}

\usepackage{arxiv}

\usepackage[utf8]{inputenc} 
\usepackage[T1]{fontenc}    
\usepackage{hyperref}       
\usepackage{url}            
\usepackage{booktabs}       
\usepackage{amsfonts}       
\usepackage{nicefrac}       
\usepackage{microtype}      
\usepackage{lipsum}
\usepackage{graphicx}
\usepackage{multirow}
\graphicspath{ {./images/} }

\title{Extraction of Energetic $N_2$ Neutrals for Efficient Plasma Food Processing}

\author{ M. Perumal$^{1}$, A. Saravanan$^{1,2}$, Prince Alex$^{1,3}$, K. V. Sunooj$^{4}$, Maheswaran Mani$^{5}$, G. Gopi Krishnan$^{5}$, \\
\textbf{Lata Shukla}$^{6}$, \textbf{P. Bharathi}$^{2}$ \textbf{and Suraj Kumar Sinha}$^{1\ast}$\\\\
 $^{1}$Department of Physics, Pondicherry University, Pondicherry-605 014, India\\
 $^{1,2}$Institute for Plasma Research, HBNI, Bhat, Gandhinagar, Gujarat -382 428, India\\
 $^{1,3}$Universita Degli Studi di Milano-Bicocca, Milan, Italy\\
$^{4}$Department of Food Science and Technology, Pondicherry University, Pondicherry-605 014, India\\
$^{5}$Department of Microbiology, Pondicherry University, Pondicherry-605 014, India\\
$^{6}$Department of Biotechnology, Pondicherry University, Pondicherry-605 014, India\\\\
  $^{\ast}$\texttt{sinhasuraj.phy@pondiuni.edu.in} \\
}

\begin{document}
\maketitle
\begin{abstract}
Low-temperature plasma (LTP) processing is a green and flexible food processing technology with only known limitation of low energy efficiency. Three established fundamental mechanisms of plasma food processing includes: i) physical damage to membranes and internal cellular components due to bombardment of energetic ions: ii) radicals and charged species led chemical reactions: iii) and radiation-induced modifications. In this investigation, we present a new mechanism of generation and extraction of energetic $N_2$ neutrals from glow discharge plasmas for efficient food processing. For nitrogen dc glow discharge plasma we demonstrate by experiment selective generation and extraction of energetic $N_2$ neutrals and $N_2^+$ ions for treatment of different food samples. Using numerical simulations, the plasma processing parameters are defined for selective extraction of $N_2$ or $N_2^+$ species with specific energies. Treatment of native finger millet flour or ragi flour (Eleusine coracana) with selectively extracted energetic $N_2$ neutrals, with an estimated average energy of 34.20 eV, resulted in improvement of shelf-life and functional properties. Additionally, this processing technique achieved high energy efficiency of $\sim$ 46\%. This method of food engineering has advantages of control over active species generation, high power efficiency, reduced processing cost, and chemical-free. Therefore, it opens up wide possibilities in plasma food processing for global food security and food safety.
\end{abstract}

\section{INTRODUCTION}
LTP (also known as cold plasma and non-equilibrium plasma) is one of the most promising food processing techniques$^{1-5}$ among the newly emerging methods of food preservation and food safety$^{6-9}$. Plasmas generated in the laboratory are only partially ionized and generally categorized as cold plasmas and hot plasmas (also known as thermal plasma).  In hot plasmas, such as welding arcs, the heavier ions and the lighter electrons are in thermal equilibrium$^{10,11}$ at temperature $\sim$10,000 K. Though species generated are highly reactive, hot plasmas are detrimental to food product quality$^{3,10}$. In LTPs, electrons and ions are not in equilibrium; the electrons are at high temperature$^{12}$ ($\sim$ 10,000 K), while the ions remain at room temperature. Therefore, LTPs do not transfer excessive heat to the sample under treatment and are considered as the least damaging to the food product under treatment$^{13}$. Three main mechanisms that dominate in a plasma treatment are: i) physical damage to membranes and internal cellular components due to the bombardment of energetic species, ii) plasma ions and radicals$^{3}$ lead to chemical reactions$^{3}$, and iii) radiation-induced damage$^{13-15}$.  Various types of LTP systems have been reported for treatment of meats$^{14,15}$, poultry$^{14}$, milk$^{14, 17}$, water$^{17}$, cereals$^{14}$, fruits$^{16}$, vegetables$^{14,16}$, as well as non-food processing using atmospheric pressure plasma$^{18}$, cold atmospheric gas-phase plasma$^{19}$, RF plasma$^{20}$, Microwave plasma$^{21}$, dc glow discharge$^{22}$, gliding arc discharge$^{23}$, dielectric barrier discharge$^{24}$, pulsed plasma$^{25}$, corona discharge$^{26}$, etc.  Though the treatment of food and non-food products with LTP exhibits superior physicochemical and functional properties, it is still at the stage of infancy$^{5}$. Accordingly, an intense investigation is required for understanding the mode of action of energetic species$^{5, 12}$, radicals$^{3}$, reactive atomic species$^{6}$, charged particles$^{12}$, radiations$^{3, 12}$, feed gas$^{3}$, applied power, energy efficacy$^{2}$, and the cost-effectiveness$^{3,4}$ of LTP food processing.\\
In plasma processing of food role of ions, electrons, and radicals is only considered though the generation of energetic neutral in glow discharge plasma is a well known phenomenon. We developed a method for the extraction of energetic neutrals for the processing of Kithul starch. A native Kithul starch sample is treated using energetic neutrals  $N_2$ extracted from air dc glow discharge plasma. The morphological damage of Kithul starch granules resulted in up-gradation of its physicochemical properties suitable for industrial application$^{7-9}$. The treated sample exhibited superior properties such as complex index (CI), enhanced starch-lauric acid complex formation, pasting property, energy storage modulus ($G'$) and energy loss modulus ($G''$) in vitro digestibility and reduction in gelatine capacities and retrogradation properties$^{7-9}$. The energetic neutral  $N_2$ molecules generated from dc glow discharge plasma could play a significant role in the treatment of food and non-food products and may find wide application in food processing.\\
In this work, we present the theoretical model for generation and extraction of energetic  $N_2$ in the range of 1.0 eV to 500 eV, followed by experimental demonstration of treatment by these neutrals on finger millet flour in the sheath region, generation of reactive atomic species, and how to achieve high energy efficacy. 
In section 2, the experimental setup is described in detail. In section 3, the method and materials are presented. In section 4, the data analysis and experimental results are presented. In the final section 5 the important findings of the present investigation is presented.
\section{EXPERIMENTAL SET-UP}
\begin{figure}[h]
\centering
\includegraphics[width=0.90\textwidth]{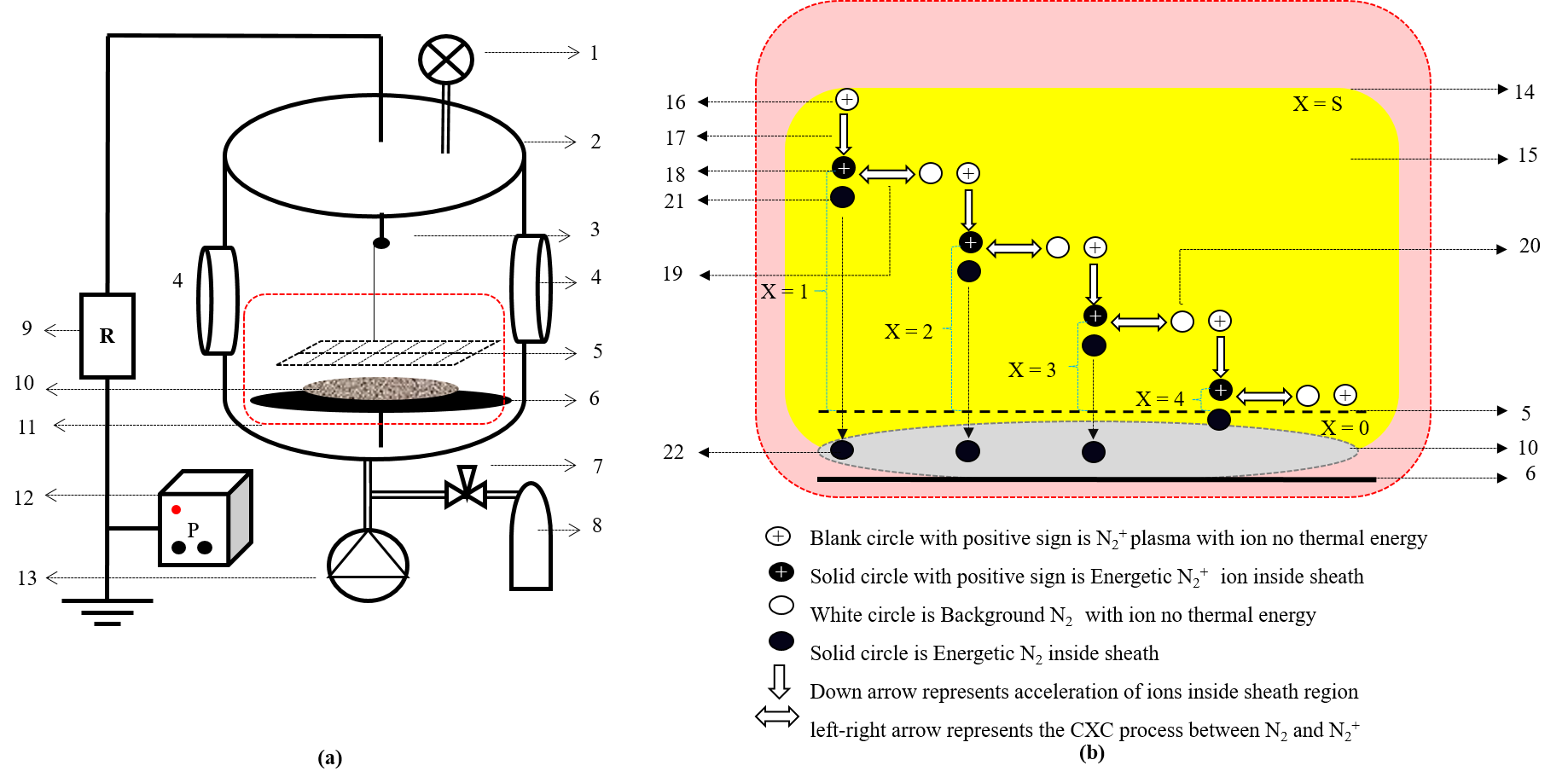}
\caption{(a) Experimental plasma system used for the treatment of starch samples. The schematics of the typical experimental condition shows different components are marked as [1] Pressure gauge, [2] vacuum chamber, [3] connection to mesh cathode [4] view port, [5] SS mesh, acting as the transparent cathode to energetic $N_2$, [6] Sample holder, [7] needle valve, [8] gas cylinder, [9] load resistor, [10] sample, [11] plasma, [12] power supply, [13] rotary pump. Fig.1 (b): Magnified view of transparent cathode region is shown: [14] is plasma cathode sheath boundary, [15] the plasma ion sheath region, [16] $N_2^+$ ion from plasma entering the sheath,  [17] acceleration of $N_2^+$ ions inside the sheath, [18] energetic $N_2^+$ ion after gaining energy, [19] left-right arrow representing the CXC process between $N_2$ and $N_2^+$, [20] the background neutral $N_2$, [21] energetic neutrals $N_2$, which is generated when energetic $N_2^+$ ion captures an electron via  CXC, [22] energetic neutrals $N_2$ striking the crossing the SS mesh electrode, which is transparent to neutral $N_2$, reaching into the bulk of the sample.} 
\label{fig:Fig_1}
\end{figure}
The experimental arrangements for the treatment of finger millet flour samples is shown in Figure. 1 (a). A negative voltage bias of -900 V is applied between the cathode and grounded chamber wall, in the pressure range of 0.2 mbar with air as operating gas. The SS mesh is shown in figure 1 (a) acts as a cathode and is connected to the negative terminal of the power supply.
Energetic $N_2$ species are generated by charge exchange collisions (CXC) inside the cathode sheath of this SS mesh. The SS mesh wire is 0.25 mm thick, and the gap between adjacent wires is 1.6 mm. The transparency of the mesh is 65\%. It is transparent to energetic $N_2$, while $N_2^+$ ions are guided by the electric field inside the sheath and strike the cathode wires and dissociate into highly active nascent nitrogen N atoms. The sample is kept below the cathode for treatment. The mechanism of the generation of energetic neutrals and its extraction is described in Figure. 1 (b). While the energetic $N_2^+$ ions dissociate into nascent atomic nitrogen on striking the cathode, provided ions have sufficient energy.  $N_2^+$ ions can also be extracted by varying the gap between mesh wires of the cathode and collisionless cathode sheath at lower pressure.
\section{METHOD and MATERIALS}
\subsection{Methods: Generation and Extraction of Energetic N2}
Understanding sheath dynamics is important for any LTP processing and critical for the selective extraction of species from the plasma. The sheath is the distance up to which an externally applied electric field can penetrate into the plasma$^{11}$. Cathode sheath (also known as ion sheath) is a positive space charge region where ions shroud a negatively biased electrode$^{12}$. An electron-sheath (also known as anode sheath) forms around a positively biased electrode$^{12}$. The cathode sheath is important for plasma processing, and we chose nitrogen discharge to investigate cathode sheath dynamics for the selective extraction of species$^{27-30}$. In a nitrogen dc glow discharge the active species observed using optical emission spectroscopy (OES) are N (742, 744, and 746 nm)$^{27, 30}$, $N_2$ (316,337,357,380, and 405 nm)$^{27, 28}$, $N_2^+$ (391, 421,470, and 522 nm)$^{27, 29}$. $N_2^+$ ions are the most dominant species, and on entering the cathode sheath, they accelerate towards the negatively biased electrodes (say, the cathode voltage $-V_0$).  If the pressure is sufficiently low, they reach the cathode surface without suffering any collisions inside the cathode sheath. For the sheath thickness(S) and mean free path ($\lambda$) for charge exchange collision (CXC) between $N_2^+$ and background $N_2$ ($\lambda$), the sheath is collisionless for condition $\lambda$ $>>$ S. For such a condition, $N_2^+$ ions accelerate inside the cathode sheath and gain energy equivalent to the applied cathode voltage (i.e., $V_0$). $N_2^+$ ions reach the cathode surface with maximum velocity ($u_m$) such that 1/2 $mu_m^2$=$eV_0$, where m is the mass of $N_2^+$. Child’s law$^{12}$ describes the potential profile inside the collisionless ion sheath. The energetic $N_2^+$ ions dissociate into nascent atomic nitrogen (N) on striking the cathode surface$^{31}$. The atomic nitrogen is highly reactive and can contribute significantly towards antimicrobial action$^{3}$. Alternatively, for relatively higher pressure, such that $\lambda$ $<<$ S, $N_2^+$ ions suffer multiple CXC inside the sheath, before striking the cathode. The energetic $N_2^+$ ions (gain energy due to sheath potential) exchange charge with background $N_2$ and become energetic neutral $N_2$. The background $N_2$, with no kinetic energy, becomes an $N_2^+$ ion, which further begins to accelerate inside the sheath. Consequently, for the collisional sheath, $N_2^+$ ions and energetic neutral $N_2$ have a velocity distribution at the cathode surface. The peak of velocity distribution is always lower than um and shifts towards lower velocity with increasing the pressure. Collisional law$^{12}$ gives the collisional sheath potential profile. In a case, few ions entering the sheath undergo CXC, and at the same time, there is a finite probability that some ion can reach the cathode without experiencing CXC. The sheath is partially collisional (0.1 $\lambda \leq S \leq \lambda$)$^{6}$ and the potential profile of the sheath, for the intermediate regime of collisionality, can be generalized in power-law form$^{6}$ as        
\begin{equation}
V_0=\frac{-V_x}{[(S-x)^b/S^b]}
\end{equation}
Where \textit{$b$} is the exponent, \textit{$x$} is the position inside the cathode sheath, $V_x$ is potential at position \textit{$x$}, and \textit{$S$} is the sheath thickness.  The value of sheath thickness S and the value of exponent \textit{$b$} is given by  
\begin{equation}
b=1/ (1-V_{0} V_{0}^{''} /(V_{0}^{'})^2)
\end{equation}
\begin{equation}
s=-bV_{0}/V_{0}^{'}
\end{equation}
\textit{$V_0$} is the electrode potential, \textit{$V_0'$} is the potential gradient, and \textit{$V_0''$} is the field gradient at the cathode. The exponent value is obtained numerically$^{6}$ using governing plasma equations$^{32}$ for a known value of discharge parameters such as pressure, applied cathode voltage, plasma density, and plasma temperature. The exponent \textit{$(b)$} approaches 5/3 for the collisional sheath and 4/3 for the collisionless sheath. Using this knowledge of the power law potential profile of cathode-sheath, the velocity distribution of $N_2^+$ ions is given by the following expression$^{6,30,32}$
\begin{equation}
        f_{i}(u,z)=2\left[\frac{us^b}{\lambda bu_m^2 z^{b-1}\left(1-\frac{u^2S^b}{u_m^2z^b}\right)^{b-1/b}}\right]exp\left[-\left(\frac{z}{\lambda}\right)\Bigg\{1-\left(1-\frac{u^2S^b}{u_m^2z^b}\right)^{1/b}\Bigg\}\right]
\end{equation}
Where z = S-x is the position of CXC inside the sheath. The energetic $N_2$ velocity distribution at the cathode, as a result of CXC, is expressed as$^{6,30,32}$
\begin{equation}
f_{n}\left(u,z=S\right)= \int_{0}^s f_{i}\left(u,z\right).\frac{dz}{\lambda}
\end{equation}
Where \textit{dz/$\lambda$} is the probability for the ion to undergo charge exchange collision with neutral within the element of \textit{dz}. The operating condition for selective extraction of the desired active plasma species is estimated numerically. For cathode voltage \textit{$V_0$} = -500 V, plasma density $n_e$ = 1x $10^{16} m^{-3}$, plasma temperature \textit{$T_e$} = 1.0 eV, and the cross-section$^{33}$ for CXC between $N_2^+$ and $N_2$ is 4x$10^{19} m^2$, the value of sheath thickness \textit{S}, mean free path, and value of exponent  \textit{$b$} obtained numerically for different pressures is listed in Table-I.   Once the value of power exponent  \textit{$b$} and sheath thickness S obtained, the velocity distribution of $N_2^+$ and $N_2$at the cathode surface is obtained numerically, using Eqn.4 and Eqn. 5$^{30}$.  At pressure 1x$10^{-4}$ mbar, {$\lambda$} = 1010.1 mm and S = 9.11 mm, and therefore $N_2^+$ ions entering the plasma sheath will reach the surface of a negatively biased electrode without suffering any collision and with energy 1/2 $mu_m^2$ $\sim$ 500 \textit{eV}. Keeping all operating parameters fixed and increasing pressures, the sheath is partially collisional, and the value of the exponent is such that \textit{b}$\approx 4/3$ for various collisionless parameters numerically estimated. Figure 2 (a) shows \textit{$f_i$} and \textit{$f_n$} due to CXC at operating pressure 0.01 mbar.  Though the sheath thickness is smaller than the mean free path (\textit{S} = 8.86 mm, \textit{$\lambda$} = 10.1 mm), there is a finite probability that $N_2^+$ ions undergo CXC before reaching the cathode. However, many reach the cathode without undergoing CXC. Therefore, the sheath is partially collisional. A sharp peak in the velocity distribution of $N_2^+$ ions at  $u_m$ = 5.80x$10^4 (m/s)$ corresponding to -500 V cathode bias voltage represents ions reaching cathode without experiencing CXC. The average energy of $N_2^+$ions and energetic $N_2$ is 396.64 eV and 360.94 eV, respectively. It appropriately indicates that $N_2^+$ ions undergo CXC near the cathode after accelerating a larger portion of the sheath. Nonetheless, the higher ion velocity than um in the graph is attributed to the energy with which these ions enter the plasma sheath boundary$^{32}$. This is the reason why the average velocity of energetic neutrals is always less than ions. As they are generated inside the cathode sheath. However, the ions with higher velocity than um are very few. The estimated average velocity of ions ($V_i$) is 5.21x$10^4 m/s$, and the average velocity of neutrals is ($V_n$) is 4.97x$10^4 m/s$. For every 100 ions entering the sheath, 125 energetic neutrals are produced by CXC at this pressure, as listed in Table-I. The increase of pressure from 0.01 to 0.05 mbar decreases the mean free path ($\lambda$) from 10.1 mm to 2.02 mm, and the sheath thickness reduces from 8.86 mm to 8.14 mm. Thus, the sheath becomes more collisional at 0.05 mbar, and accordingly, the peak of  \textit{$f_i$} and \textit{$f_n$} shifts towards lower velocity.
\begin{figure}[h]
\centering
\includegraphics[width=0.90\textwidth]{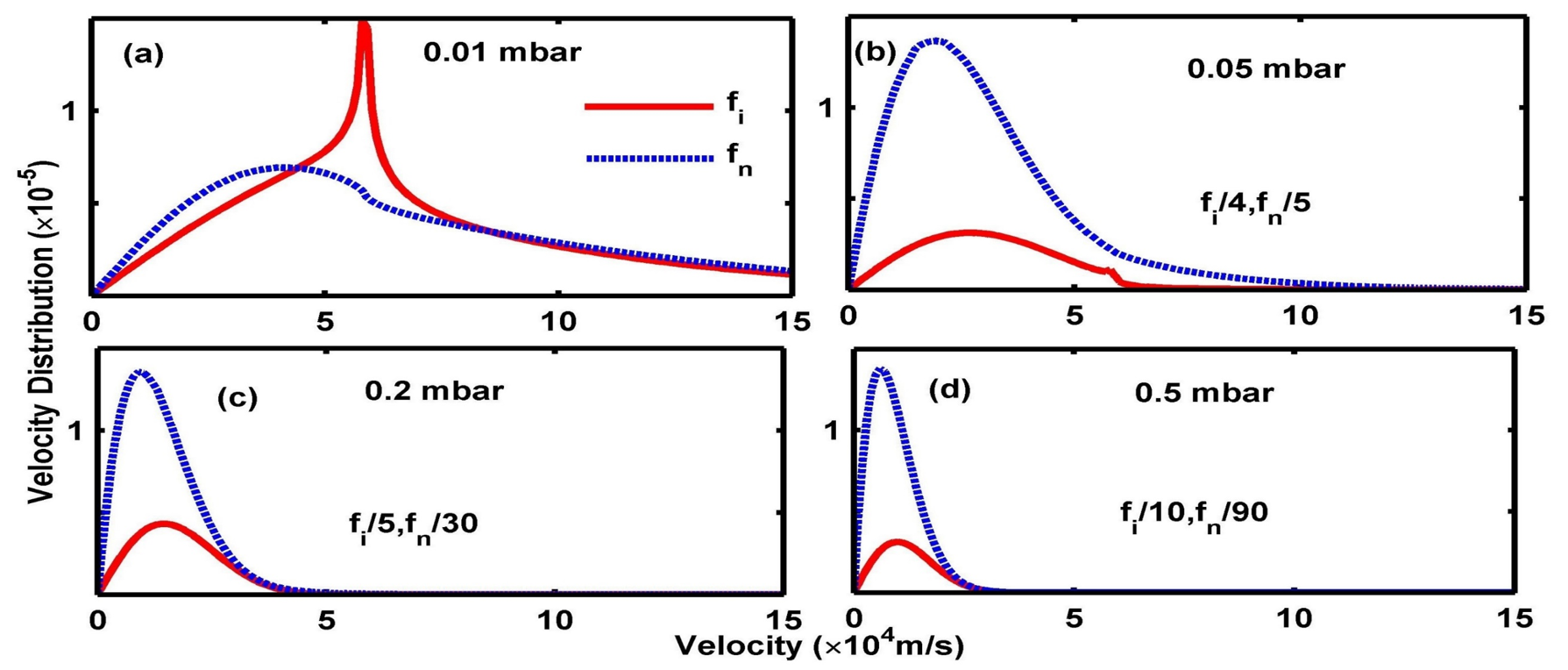}
\caption{shows \textit{$f_i$} and \textit{$f_n$} distributions for fixed cathode voltage at -500V and with the increase of pressures (a) 0.01 mbar, (b) 0.05 mbar, for this case, as the \textit{$f_n$} increases rapidly, for better visualization and its comparison at different pressures, axes scale are fixed and fi are divided by factor of 4 and  \textit{$f_n$} are divided by factor of 5, (c) 0.2 mbar \textit{$f_i$} divided by a factor of 5 and  \textit{$f_n$} divided by a factor of 30, (d) 0.5 mbar,   \textit{$f_i$} divided by 10 and \textit{$f_n$} divided by 90, with the increase of pressure velocity peak shifts towards lower velocity and neutral peak increases.}
\label{fig:Fig_2}
\end{figure}
For this case, every 100 ions entering the sheath generate nearly 530 energetic neutrals on their way to the cathode surface as shown in figure 2 (b). With the fixed scale of the x-axis (x $10^4 m/s$) and y-axis fixed (x $10^{-5}$) and for better visualization, \textit{$f_i$} is divided by factor 4 and \textit{$f_n$} divided by a factor of 5 for this pressure. Figure 2. (c) and 2 (d) shows \textit{$f_n$} and \textit{$f_i$} at pressure 0.2 and 0.5 mbar, respectively. The same trend is maintained and the number of energetic neutrals generated per ion increases, and the peak of \textit{$f_i$} and \textit{$f_n$} further shifts towards lower velocity. For pressure 0.2 mbar and 0.5 mbar, \textit{$f_i$} is divided by a factor of 5 and 10, and \textit{$f_n$} is divided by a factor of 30 and 90 respectively, for the same reason as mentioned above. The energy of ions and neutrals reaching the cathode decreases with increasing pressures and is listed in Table-I. At 0.2 mbar, the energy of $N_2^+$ ions and energetic $N_2$ reach cathode with energy $\sim$ $45$ eV, and $\sim$ $29$ eV. Under relatively higher pressure conditions, such as 10 mbar, cathode voltage = -500 V, the mean free path is $\lambda$ = 1.01 x$10^{-2}$ mm and S = 3.38 mm. The energy of $N_2^+$ and energetic $N_2$ neutrals at the cathode surface is as low as $\sim$ $2.0$ eV.
\begin{center}
\begin{table}
	\centering
	\caption{Different plasma parameters estimated at $V_0$ = -500 V and $n_e$ = 1x$10^{16}$ $m^{-3}$. \textbf{Condition for treatment of native Kithul starch.}}
		\begin{tabular}{|c|c|c|c|c|c|c|c|c|c|}
	\hline
		 &         &       &     &Power    &\multicolumn{2}{c|}{Ion}&\multicolumn{2}{c|}{Neutral}&No. of $N_2$\\
		\cline{6-9}
		  & P       & $\lambda_{mfp}$      & $S$ &exponent &&&&&generated\\
		V  & (mbar)  & (mm)   &(mm)       &b value  &&&&&  per ion\\
		&   &   &   &   & $V_i$ (x$10^4$) & Energy  & $V_n$(x$10^4$) & Energy &   \\
		&   &   &   &   & (m/s)  & (eV)    & (m/s)  &  (eV)  & \\
		
		\hline 
		
		\multirow{6}{*}{-500}
		
		& 0.0001 & 1010.10 & 9.11 & 1.34 & 5.84 & 500 & -- & -- & -- \\
		
		\cline{2-10} 
		
		& 0.01 & 10.1 & 8.86 & 1.41 & 5.21 & 396.64 & 4.97 & 360.94 & 1.25 \\
		
		\cline{2-10} 
		
		& 0.05 & 2.02 & 8.14 & 1.55 & 3.18 & 147.76 & 3.05 & 135.93 & 5.30 \\
		
		\cline{2-10} 
		
		& 0.1 & 1.01 & 7.57 & 1.60 & 2.32 & 78.65 & 2.04 & 60.81 & 9.14 \\
		
		\cline{2-10} 
		
		& \textbf{0.2} & \textbf{0.50} & \textbf{6.90} & \textbf{1.63} & \textbf{1.76} & \textbf{45.26} & \textbf{1.41} & \textbf{29.05} & \textbf{15.57} \\
		
		\cline{2-10} 
		
		& 0.5 & 0.20 & 5.96 & 1.65 & 1.21 & 21.39 & 0.92 & 12.36 & 31.49 \\
		
		\hline
		\end{tabular}
		\\{V = Voltage, P = Pressure, $\lambda_{mfp}$ = Mean free path and \textit{S} = Sheath thickness}.
\end{table}
\end{center}
This numerical estimation set the basis for the selective extraction of active species. High energy $N_2^+$ ions can be extracted using a gridded cathode with high transparency (70\%-90\%) for operating pressure such that  \textit{$\lambda$ $>>$ S} for example at pressure  1x$10^{-4}$ mbar, discussed above.  The separation between wires of the gridded cathode is such that it provides no obstructions to ions$^{28}$. On reducing the transparency of gridded cathode up to 20\%, for this case, most of the ions will strike the cathode grid wires with energy ~ 500 eV and likely dissociate into highly reactive nascent atomic nitrogen N. Energetic $N_2$ neutrals are selectively extracted when the operating pressure is such that sheath collisionality is in intermediate regime (0.1 $\lambda$ $\leq$ S $\leq$ $\lambda$) and transparency of gridded cathode is 50\%-70\%. For example- at pressure 0.2 mbar for the present case, as nearly 16 energetic $N_2$ neutrals are generated per $N_2^+$ ion. These ions have a high chance to follow the electric field and strike the grid wires of the cathode and therefore under the typical condition probability of $N_2^+$ reaching the substrate is minimal. Further, at relatively high pressures such as 10 mbar, $N_2^+$ and energetic $N_2$ neutrals have significantly low energies, and for the low grid, transparency will result in intense characteristic radiation of the operating gas.  We investigated nitrogen plasma for its merit of high CXC leading to control over the generation of active species it is important to mention that active species generation can be further altered by addition of a small fraction of $CH_3$, $H_2$, Ar, $O_2$, He, Xe in operating gas can generate additional active species such as ions$^{27}$ ($H3^+$, $NH_3^+$, $NH_4^+$, $NH_2^+$, $N_2H^+$, $NH^+$, $H_2^+$, $H^+$, $Ar^+$, $ArH^+$, $N^+$), radicals$^{3, 27}$ (H, N, NH, $NH_2$, $NO_x$, $O_3$) and the characteristic radiations. 
\subsection{Materials}
Pulverized finger millet flour is used as an edible nutrients in daily life in many localities. The sample collection finger millet flour used for the present study was procured from the local market. The sample was tightly packed and maintained at room temperature.
\subsection{Characterization Techniques}
\subsubsection{X-ray diffraction study}
The crystalline structure of native and plasma treated Finger millet flour is identified using an X-ray diffraction (XRD) study in PANalytical Xpert PRO diffractometer. The diffraction pattern recorded using Cu-$K\alpha_1$ radiation with the wavelength of 1.5406\AA and the instrument is operated at 40 kV and 30 mA. The diffraction pattern is recorded with the scanning angle (2$\theta$) range of $20-18^{\circ}$ at the step size of $0.02^{\circ}$.
\subsubsection{FTIR Analysis}
Fourier Transform Infrared (FTIR) spectra of native and treated samples were determined using a Shimadzu-8700 FTIR spectrometer over the scanning range of 4000 - 400 $cm^{-1}$ at room temperature and a scan rate of 20 per sec. The samples were mixed with laboratory grade KBr ($1:100$) and made into pellet by pelletizer for the IR analysis.
\subsubsection{Retrogradation tendency}
The retrogradation tendency method described in the tendency of plasma treated finger millet flourss measures the percentage of light transmittance at 650 nm against the distilled water blank using a Shimadzu UV-VIS 1700 PharmaSpec spectrophotometer. The flour sample (0.15 g) was mixed in 15 ml of distilled water using a 3 incubate of the slurry in a water bath at $1000 ^{\circ}$C for 30 minutes with intermittent stirring. Incubate the slurry was cooled at room temperature for 1 hr and the percentage of light transmittance has been measured. In this analysis to monitor the retrogradation tendency, we store the slurry at 40 $^{\circ}$C for 7 days.
\subsubsection{Syneresis}
Syneresis (freeze thaw stability) of samples was a mixture of 0.6 g of plasma treated and native flour samples in 10 ml of distilled water. After freezed slurry samples was maintained at - $18^{\circ}$C for 22 hours. Thawing the slurry in a water bath at 30$^{\circ}$C for 90 minutes, then the slurry was centrifuged (HERMLE Z300 K) at 1180 x g for 15 minutes, after transfer the supernatant into a beaker and the residue weigh the supernatant. The percentage of syneresis was then calculated as the ratio of the liquid decanted weight and the total weight of the gel before centrifugation multiplied by 100. The syneresis was measured four consecutive days freeze-thaw cycles were performed.
\section{Result and Discussion}
We treated finger millet flour (Eleusine coracana) in nitrogen glow discharge plasma at pressure 0.2 mbar and cathode bias -900 V. Under this operating condition, the sheath thickness is S = 10.0 mm; the mean free path of  $N_2^+$ is $\lambda$ = 0.505 mm. Each  $N_2^+$ ions generate 21.77 energetic  $N_2$ neutrals. The numerically estimated average energy of  $N_2^+$ ions and energetic  $N_2$ neutrals is 55.56 eV and 34.20 eV, respectively. The mesh cathode transparency is $\sim$ 65\% for energetic  $N_2$, the thickness of mesh wires is 0.25 mm, and the gap between adjacent wires is 1.6 mm. For this case, dominant active species are energetic $N_2$ neutrals. The characterization of treated Finger millet floursamples shown in Fig. 3. Fig.3 (a) shows the X-ray diffraction patterns observed for the native and the plasma-treated sample. The main diffraction peaks of Finger millet flour are observed at angles  $15^{\circ}$, $16^{\circ}$, $18^{\circ}$, and $23^{\circ}$. The intensity of the peaks depends on the number of ordered semi-crystalline structures in the flour. The plasma treatment has no effects on its crystal structure, though the relative crystallinity of the treated and untreated sample differs. The infrared transmittance spectrum of the Finger millet flouris depicted in Fig.3 (b). It exhibited three IR peaks at 3469 $cm^{-1}$, 2629 $cm^{-1}$, and 1647 $cm^{-1}$, showing the O-H stretching vibrations, $CH_2$ stretching vibrations, and H-O-H bending vibrations, respectively$^{34}$. The intensity of these absorption peaks increases on plasma treatment compared to native Finger millet flour, and no chemical groups were created during the plasma treatment. The increase of the peak sharpness corresponding to $CH_2$ and O-H stretching vibrations and H-O-H bending vibrations is mainly due to the breakage of hydrogen bonds$^{34}$ by the energetic $N_2$ to flour samples. Retrogradation analysis$^{35}$ is shown in Fig.3 (c) also indicates that treated flour exhibited increased long-term storage compared to that of untreated flour during the incubation period of 7 days. An increase in syneresis$^{36, 37}$ is shown in Fig.3 (d) the depolymerization of starch molecules are induced by the glycoside bond's fission. We conclude that the plasma-treated flour exhibit favourable properties for wide perishable food treatment and breaking down of complex polypeptide, amylopectin, and also phosphodiester linkages applications. The treatment also shows an increase in solubility and swelling capacity$^{35}$ of the treated flour. 
\begin{figure*}[h]
    \centering
    \includegraphics[width=0.45\textwidth]{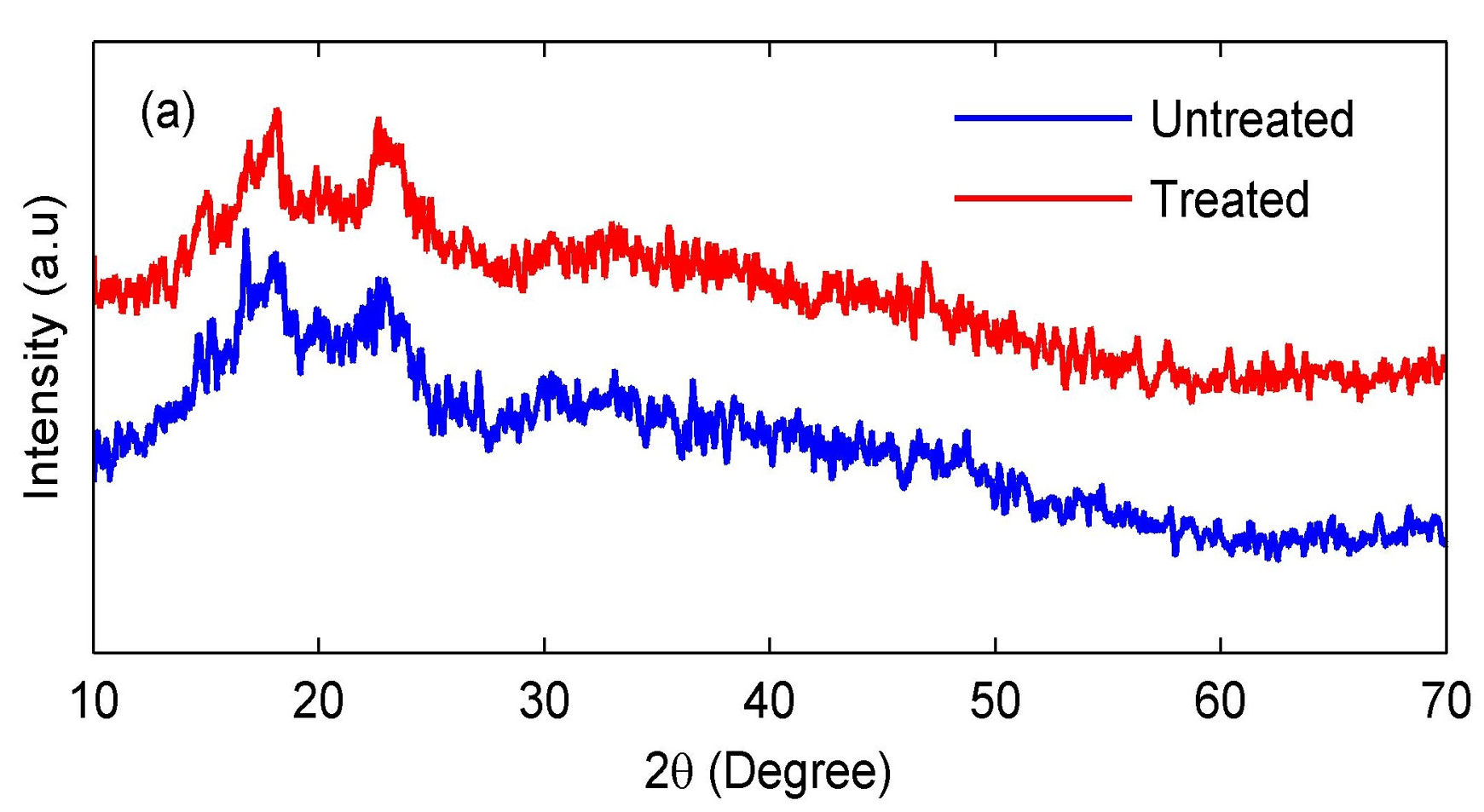}
    \includegraphics[width=0.45\textwidth]{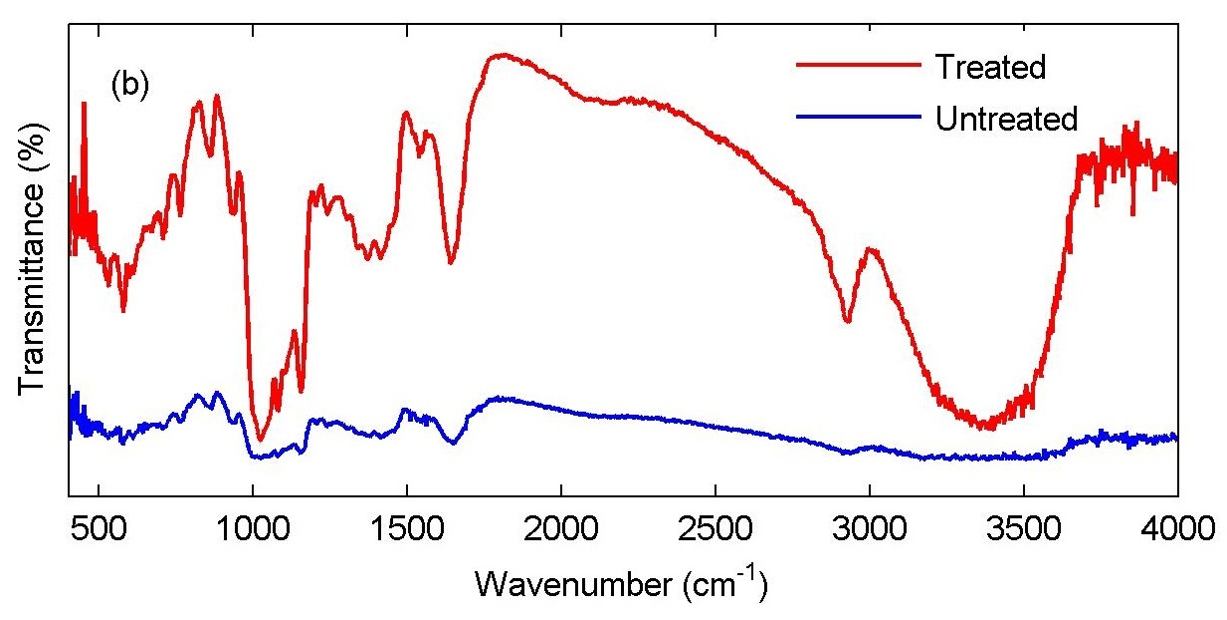}\\
    \includegraphics[width=0.45\textwidth]{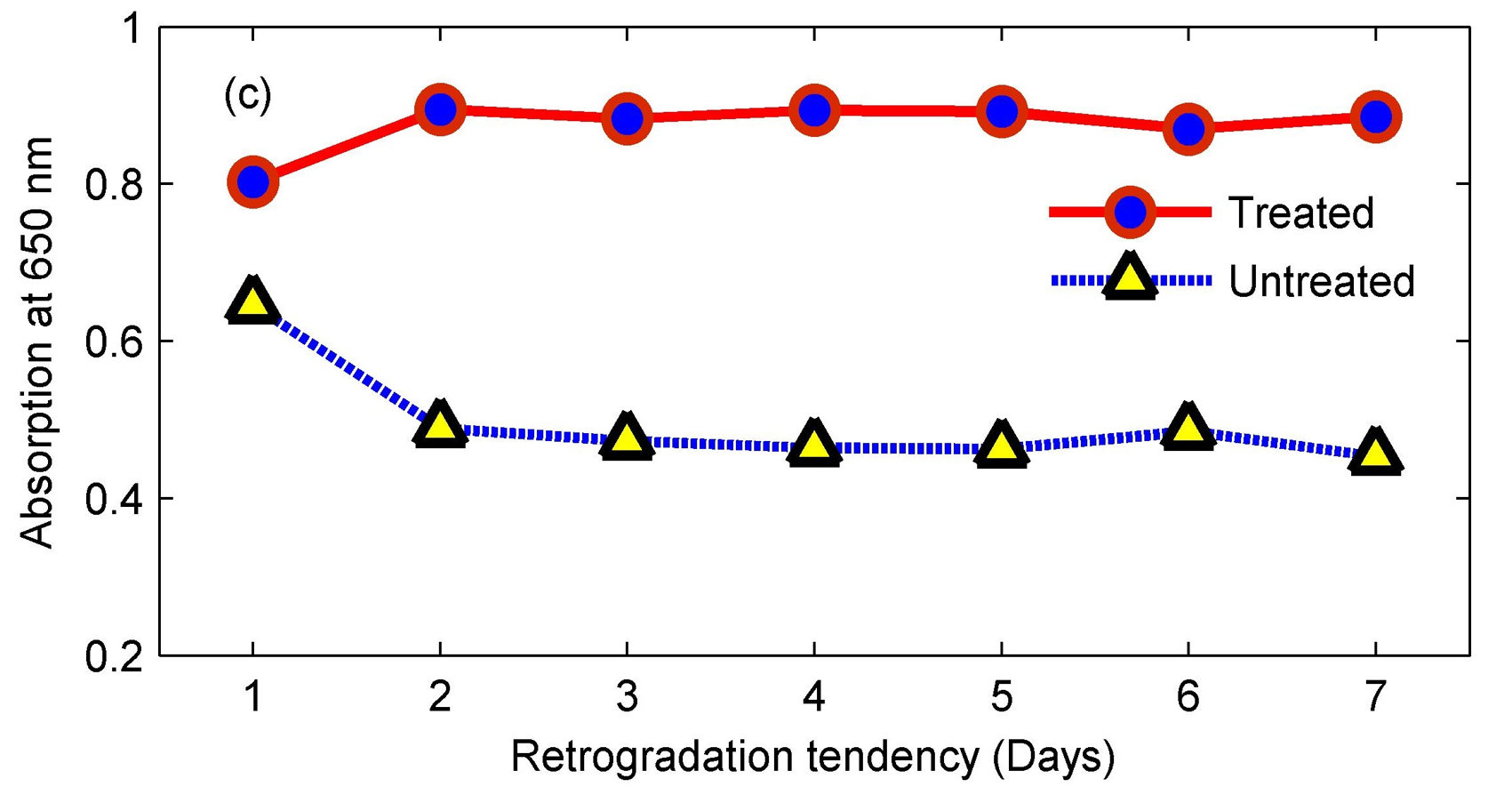}
    \includegraphics[width=0.45\textwidth]{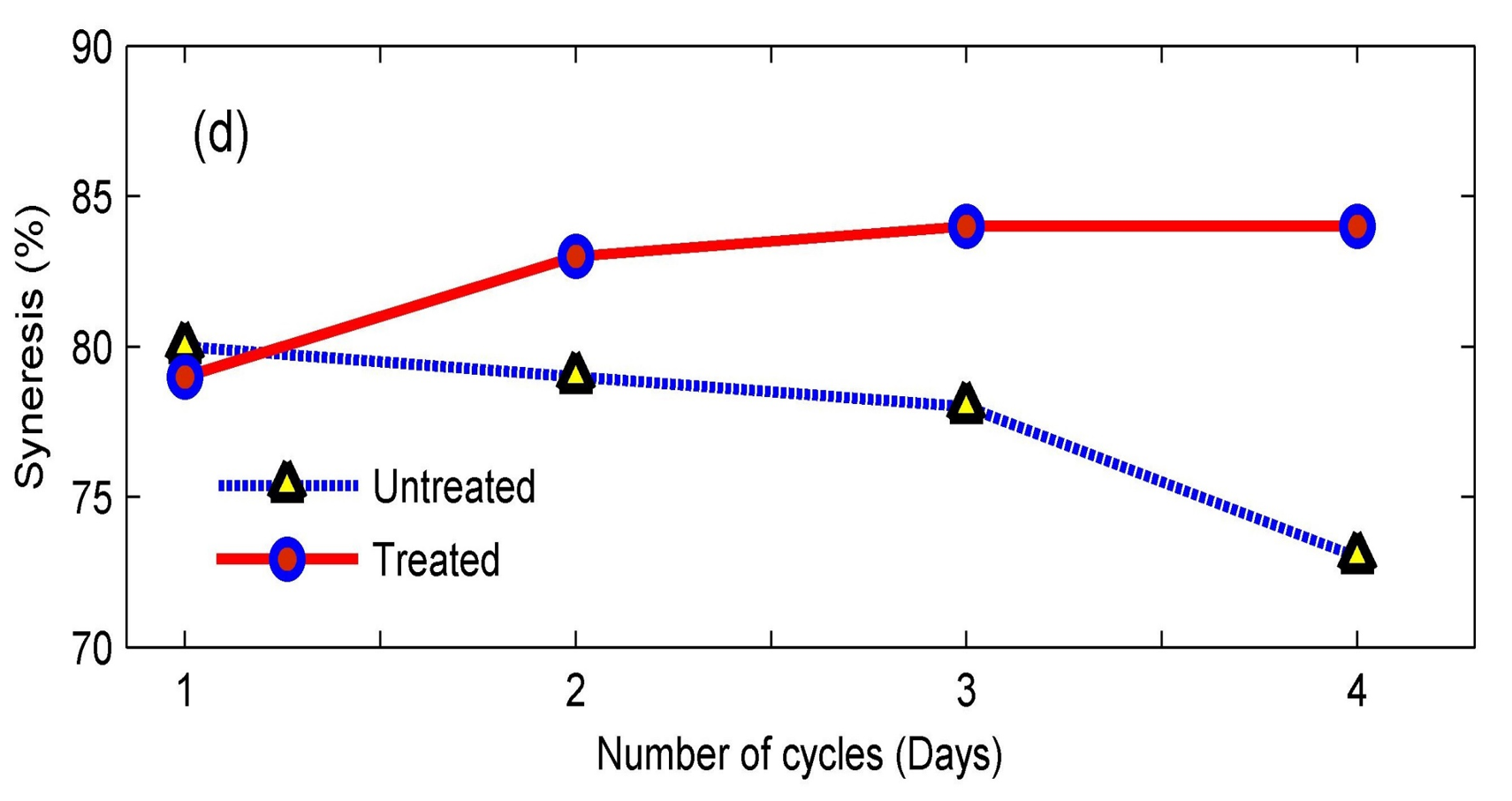}
    \caption{(a) X-ray diffraction (b) FTIR spectra analysis (c) Retrogradation analysis (d) Syneresis analysis for the Finger millet flour.} 
   \label{fig:Fig_3}
\end{figure*} \\
In addition to the merits of the LTP process, the economics of treatment is an essential factor in its commercial success, and for food processing, low cost is the desired goal to achieve global food security. Therefore, we focussed on two fronts: i) reduction of feed gas and ii) energy efficiency of the process. The cost of nitrogen as feed gas estimated is $ 9,000 - $72,000; for 1,000 hours of operation$^{3}$. In our previous work, we tested the applicability of the process with air as a feed gas for native Kithul starch sample treated using energetic neutral $N_2$ extracted from air dc glow discharge plasma at 0.2 mbar and cathode voltage -500 V. Air contains 80\% nitrogen, and it can eliminate this cost, with the additional advantage of $O_3$ and nascent atomic nitrogen (N) production$^{3}$, which is useful for antimicrobial action$^{3}$. It showed the up-gradation of its properties suitable for industrial application$^{7-9}$. The morphological damage of Kithul starch granules resulted in up-gradation of its physicochemical properties$^{8}$. In the present treatment with nitrogen dc glow discharge we estimated the energy efficient of the process. However, some of these energetic $N_2$ neutrals are obstructed due to the limited transparency (T) of the mesh electrode and the total discharge power ($P_T$). Thus, the fraction of electrical power ($P_N$) carried energetic $N_2$ utilized in the process can be approximated as
\begin{equation}
P_{N}= \frac{P_T}{\ (1+ \gamma_E)} X T
\end{equation}
Therefore, the power efficiency of the process ($\eta$) is expressed as
\begin{equation}
\eta = \frac{P_N}{\ P_T} = \frac{[\frac{P_T}{\ (1+ \gamma_E)} X T]}{\ P_T} = \frac{T}{\ (1+ \gamma_E)} X 100
\end{equation}\\
Therefore, the power efficiency of the process is defined by the transparency of the cathode and the effective secondary electron emission coefficient $\gamma_E$ (ESEEC) in dc glow discharge conditions. The transparency (T) of the mesh cathode is 65\% for our case, and accordingly, the efficiency is in the range of 48\% - 45\% for $\gamma_E$ in the range of 0.30 - 0.38. The generation of the energetic $N_2$ neutral as active species enhanced the energy efficacy of the process. Under the typical operating condition, 80\% of the total discharge power goes to $N_2^+$ ions and $\sim$ 90\% of this energy further distributed to energetic $N_2$ neutrals. Thus, energetic $N_2$ neutrals carry 	$\sim$ 72\% of the total electrical power. Again 35\% of these energetic $N_2$ neutrals obstructed at gridded. Therefore, 	$\sim$ 46\% of the total electrical power is dumped to the sample under treatment. The controlled generation and selective extraction of active species from LTP is applicable to a wide range of food and non-food processing for value addition.
\section{Conclusion}
From the present investigation, we draw the following major conclusions:\\
i) Using the present technique, the glow discharge plasma based food processing can be applied over a large variety of products, as the required active species such as energetic ion, neutral, radicals and radiations can be selectively extracted as per the desired modifications.\\
ii)	Flux of energetic $N_2$ and $N_2^+$ ions can be controlled and active species with specific energy can be extracted for the treatment. In the present experiment condition, the numerically estimated average energy of $N_2^+$ ions and energetic $N_2$ neutrals is 55.56 eV and 34.20 eV, and $N_2$ flux (energetic neutral) is 16 time of $N_2^+$ flux (energetic ion).\\  
iii) Under the typical conditions, 80\% of the total estimated discharge power goes to energetic $N_2$ neutrals, and after extraction $\sim$ 46\% of discharge power is dumped to the sample under treatment, which is considerably high for plasma processing.\\
iv)	The processing gas cost can be eliminated by using air as a replacement of $N_2$ gas since air contains $\sim$ 78\% $N_2$.\\
v) The plasma treated finger millet flour exhibited superior properties such as enhanced storage time, increases molecular depolymerization of syneresis, functional properties like solubility, swelling increased after plasma treatment and physicochemical properties.\\ 
The present investigation highlights the possibility of cold plasma food processing for wide commercial applications.
\section{Acknowledgements}
We acknowledge the scientific and technical assistance of Prof. S. Mukherjee, Institute for Plasma Research, HBNI, Bhat, Gandhinagar, Gujarat 382 428, in developing the ‘sheath theory’ for neutral energy distribution and the analysis of the experimental results. This work is partially funded by the University Grant Commission (UGC), India, under the project F.No.41-970/2012 (SR) and the Department of Science and Technology (DST), India, under the project SR/FRT-PS-053/2010. We also thank Pondicherry University for the Start-up Grant.
\section{Author Contribution}
SKS overall research supervision, conceptualization, visualization, innovations, data curation, validation, writing, theoretical and numerical investigation, review, editing. MP carried out the experiment, data curation and analysed data. MP and AS numerical estimations, and original draft-writing. PA design the vacuum system, installation and calibration, and editing. KVS, MM, LS and GG formal analysis and writing. PB contributed to the spectroscopic identification of species in nitrogen plasma and editing. All authors reviewed and approved the final version of the manuscript.



\end{document}